# Performance Study and Simulation of an Anycast Protocol for Wireless Mobile Ad Hoc Networks

Reza Azizi

Engineering Department, Bojnourd Branch, Islamic Azad University, Bojnourd, Iran

reza.azizi@bojnourdiau.ac.ir

## ABSTRACT

*This paper conducts a detailed simulation study of stateless anycast routing in a mobile wireless ad hoc network. The model covers all the fundamental aspects of such networks with a routing mechanism using a scheme of orientation-dependent inter-node communication links. The simulation system Winsim is used which explicitly represents parallelism of events and processes in the network. The purpose of these simulations is to investigate the effect of node's maximum speed, and different TTL over the network performance under two different scenarios. Simulation study investigates five practically important performance metrics of a wireless mobile ad hoc network and shows the dependence of this metrics on the transmission radius, link availability, and maximal possible node speed.*

## 1. INTRODUCTION

Any computer network, which is not connected by the cables and in which data is transmitted by using radio waves between nodes of the network is called a wireless network. Wireless networks support mobility, so users have access to network anywhere within the range. Also installing a wireless network is simpler and faster due to the elimination of cables.

Wireless ad hoc network [1] is a type of wireless network that does not need any existing infrastructure such as wireless router or access point. An ad hoc network consists of multiple nodes that are connected through wireless links. Since the transmission range for each node is limited, if receiver node is not inside the coverage area of sender, each node should participate in routing as intermediate node by forwarding data to other nodes until it reaches the receiver. It means every node can work as a router in network to establish a multi-hop wireless link between sender and receiver. A mobile ad hoc network [2] (MANET) is a decentralized wireless ad hoc network in which nodes can move arbitrarily in any direction; therefore it results in frequent changing of links to other nodes. Like in other wireless ad hoc networks, every node should forward the data which is not related to it, and accordingly act as a router. Due to dynamic nature of MANETs, they have wide usage in military scenarios to disaster relief operations or sensor networks. Besides, they are also used increasingly in our everyday life for transferring the data between wireless devices, and mainly to share internet in home networks or public places like airports, restaurants.

As mentioned above each node must be able to work as a source, destination, or router and decide which way to route packets. The act of selecting paths to direct the packets or generally network traffic is called routing. Routing protocol is the tool used to control all the transmissions inside the network. It also should be able to handle the topology changes as a result of node's random movement through the wireless network. Flooding [3] is an algorithm in which every incoming message is sent to all reachable parts of the network. It is easy to implement and is used as a part of some routing protocols. Besides, anycasting algorithm is used to choose the topologically nearest node in a group of possible receivers and forward data toward it.

Simulation modeling [8] and real-world experiments [4] are two different methods used for investigation and performance evaluation of wireless ad hoc networks. The simulation method allows studying the characteristics and manner of MANETs with eligible number of nodes and desired number of repetitions, in comparison with real-world investigation that requires much more resource and time and to provide precise and noteworthy information.

A detailed simulation study of stateless anycast routing in a mobile wireless ad hoc network is conducted. The proposed scheme enables representation of reliability aspects of wireless communication in a general and flexible way. Using a flooding anycast mechanism, the paper addresses issue of locating the nearest server from a group of contents-equivalent servers in the network. The simulation model explicitly represents parallelism of events and processes in the wireless network. The goal of this study is to investigate an anycast routing protocol characteristics in wireless ad hoc network under different conditions with use of some performance metrics. In simulation, the behavior of five fundamental performance metrics - response ratio, average number of hops, relative network traffic, average response time and duplicate ratio - was investigated with varying distance of transmission and different combinations of model parameters.

The rest of the paper is organized in the following way. Section 2 presents system assumptions and specification of the chosen mobility model in simulation modeling. Section 3 explains the simulation setup, organization of our conducted simulations, and performance metrics. Respectively, sections 4 and 5 mention results and their discussions. Finally Section 6 concludes the study.

## 2. SIMULATION SYSTEM

Winsim [5] is the simulation system which used for modeling and simulation of wireless ad hoc network in this paper. Winsim is based on a class of extended Petri nets. It has high level of programming language possibilities for processing complex data, and provides quick simulation. More details and examples about extended Petri nets are provided in [6]. The program was developed based on the prototype program already implemented in [7].

### 2.1. The System Assumptions

The area of network is assumed to be restricted to a rectangular shape with system configuration parameters $x_{min}$ and $x_{max}$ for horizontal axis, and $y_{min}$ and $y_{max}$ for vertical axis. Also the number of the nodes within this area is fixed and their primary distribution assumed to be random with uniform probability distribution within the ($x_{min}$ , $x_{max}$ ) and ($y_{min}$ , $y_{max}$ ) limited areas.

Next assumption is that, nodes have capability to communicate with each other, by using of bidirectional wireless channels. The transmission radius is assumed to be same in different directions. Besides, even within this limited coverage area, inter-node connection is not reliable due to different reliability aspects of wireless communication, like interference, fading, or climate conditions. Each node has a unique identifier or address.

Another assumption is that, movement of the nodes in the given area is same in form with a chosen mobility model. A node will bounce and continue moving within the area in a new direction, if it reaches to the borders.

This model assumes that nodes change position alternatively at discrete steps. This time interval for each step, is another system configuration parameter, and is defined by τ. Also each node is moving with a different speed in the range of $(0, \Delta V_{max(x)})$ and $(0, \Delta V_{max(y)})$.

## 2.2. Mobility Model

A mobility model controls the movement of mobile nodes, and change in their speed and location. The Random direction mobility model [10] is one of commonly used synthetic models. In this study, the random direction mobility model is used, but with some modifications. In the modified version, mobile nodes still continue to select random directions but can change their direction of movement at the end of any step, with the probability $p$. So nodes are not forced anymore to reach the borders to choose a new direction. By using the $p$ probability as another system configuration, we can demonstrate various motion patterns. The original random direction model can be reached, if value of $p$ is set to zero in extended version.

A flooding-based [11] simulation system was developed relying on the chosen mobility model. System is used to localize an anycast server in wireless ad hoc network by employing anycast service. It is assumed that there are two types of nodes in the network. Simple nodes (clients) are the first type of nodes, they are sources of anycast requests. Simple nodes (intermediate) re-transmit anycast requests, which come from source nodes, in multicast mode inside the network area. Simple nodes are also capable of forwarding unicast replies generated by server nodes.

It is assumed that there is one group of anycast servers in the network with five identical mobile server nodes. These server nodes are distributed randomly in the network area. In the current model, there is only one source of requests. Therefore, logically, the system is equivalent to a finite population queuing system [12] with one client and a few identical servers.

## 3. SIMULATION SETUP

Simulation experiments were organized and conducted according to the following setup. It is assumed that the network area is a rectangular (square) of 500 m × 500 m. Such an area is quite realistic for small and medium-sized ad hoc wireless networks. The network area is populated by $N=50$ nodes, having numbers $1, 2, 3,..., N$. The first m nodes are anycast servers with numbers $1, 2, ..., m < N$. The number of anycast servers, m, is specified in the file of parameters. The nodes with numbers $m+1, m+2, ..., N$ are simple nodes. It is assumed that $m<N-m$, i.e. the number of anycast server nodes is less than the number of simple nodes. For the sake of simplicity, simple node N is the source node that generates anycast requests.

Initial positions of the nodes (simple and servers) are random and different in different simulation runs with the uniform probability distribution [6] in given area. That is, the network area with its nodes can be approximated as a point Poisson field [6]. All nodes move from their initial positions according to the chosen mobility model.

Any anycast server can receive more than one multicast request, but only the first received multicast request will be accepted and responded. Any simple node can receive more than one unicast reply from a few servers, but only the first unicast reply will be accepted and forwarded (if not the source node).

As was mentioned earlier, only one network node was used as a source of anycast request messages. All other network nodes work as message routers or servers of anycast messages transmitted by the source node. Correspondingly, the source node discards all requests that can be transmitted by other nodes since these requests are copies of messages initiated by the source node. The source node assigns a unique number to each generated packet. Interval between transmissions of requests by the source node is set to be 500ms. For a small sized or medium sized ad hoc network, this interval is sufficiently large to complete all activities in a network related to a request transmitted by the source node before it transmits the next request. As a result, at any moment of simulation time, the model will handle, at different nodes, messages with the same identifier. This considerably simplifies the model and its study.

To obtain sufficiently stable statistical results of simulation, the total number of requests transmitted by the source node is set as Ns = 2000 messages. With this number of messages and inter-message interval of 500 ms, the simulation interval of each run is 2000 × 500 ms + 1000 ms = 1001000 ms, where 1000 ms is a small margin to provide the clearance of the model at the end of each simulation run [9].

Starting from a chosen random position, each network node (including the source node) moves in a random direction with a constant random speed in the given area. The random speed of a node is set according to uniform probability distribution in the range $(0, V_{max})$, where $V_{max}$ is the maximal speed set as a network configuration parameter.

As it is explained in [13], the inter-node communication is considered as very reliable for nodes, which are very close to each other. For this reason, in the model, the distance to very close nodes is assumed to be a random variable which has a lower bound equal to zero and upper bound being uniformly distributed from 5 to 10 meters. Also when probability of changing direction is equal to zero, nodes change the direction of their movement randomly at the border of the network area was used in this mobility model.

One more parameter of the inter-node communication scheme is the interval in which the states of oriented inter-node communication links are checked. This interval is the same for all simulation experiments and was set at 2000ms. At the end of this interval, the state of each link can change.

The simulation experiments were conducted for maximal transmission distances 30m, 60m, 90m, 120m, 150m, 180m and 210m. Obviously, with these distances, the message transmitted or forwarded by a node can reach only a subset of network nodes in the given area. This is true for real ad hoc wireless networks.

It is also assumed that each network node, intermediate or destination one, can lose any message, transmitted by another node, with some probability l. In the simulation experiments, as parameters, link availability $l$ was used as the message loss probability in the range of $0 < l < 1$. The value of time-to-leave (TTL) field in generated packets was fixed at seven or four in each request message.

Two series of experiments were conducted. In the first series, the chosen performance metrics were studied for maximal node speeds 5 Km/h, 30 Km/h and 50 Km/h. It should be noted that, with the given value of $V_{max}$, different network nodes will move with different speeds in the range $(0, V_{max})$. For these series of experiments link availability $l=0.7$ and probability of changing direction $p=0.0$ were used. In both series of experiments, the value of TTL in each request message was fixed at seven.

In the second series of conducted experiments, the effect of TTL value on the performance metrics was investigated. For this reason, a set of experiments were performed by setting TTL to 4 and 7 with link availability $l$ = 0.05, 0.1, 0.3, 0.5 and 0.7 when $V_{max}$ was set to 5 Km/h, probability of changing directions $p=0.0$ and the maximum transmission distance is varied in the range (30-210) meters. All the parameters and setup of simulation setup are shown in Table 1.

Table 1: Parameters of simulation setup

| | |
|---|---|
| Network area | 500 m x 500 m |
| Number of nodes | 50 |
| Total number of requests | 2000 |
| Interval between transmission of requests | 500ms |
| TTL (Time-to-leave) | 4 and 7 |
| Link availability ($l$) | 0< $l$ <1 |
| Maximal transmission distances, m | 30 to 210 |
| Maximal node speed ($V_{max}$) | 5 Km/h, 30 Km/h and 50 Km/h |
| Changing direction probability ($p$) | $p = 0$ |

### 3.1. Performance Metrics

Performance metrics are used to help researchers to investigate the wireless ad hoc networks. Delivery ratio and average number of hops per delivered packet are the most popular performance metrics used for this reason. The delivery ratio (referred as response ratio) characterizes how the network is effective in delivering packets from source nodes to server (destination) nodes. Average number of nodes that a packet traverses in its way to the source node is represented as the average number of hops. Both of these performance metrics have direct relation with the implemented routing algorithm, node mobility models and inter-node communication links specification.

Response time is another performance metric. Response time is the time interval between the moments the source node sends a request, until the reception of the reply message. This metric is important for some real-time applications which need small time interval.

Each packet that is transmitted by a source node will be usually retransmitted by some intermediate nodes until it is received by server nodes. Relative traffic is the performance metric that represents the number of times each packet is transmitted by other nodes. As matter of fact, it is necessary to keep relative traffic as low as possible to have less overloading in the network.

All received replies in source node for a request, after receiving the first reply are taken as duplicated replies. This characteristic of the network behavior is shown by duplication ratio performance metric. It has a direct relation with robustness and availability of the network, but should not be large to have less traffic in the network.

## 4. RESULTS OF SIMULATIONS

The results of the first series of simulations are presented in Figures 1 – 5.  The results of the second series of simulation are shown in Figures 6 – 15. Afterwards, Figures 16 – 20 represent the comparison for second series of experiments.

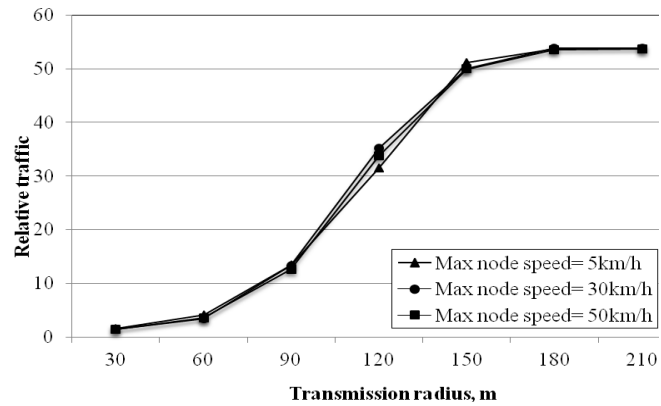

Figure 1: Relative traffic versus transmission radius with link availability *l*=0.7 and different maximal node speed.

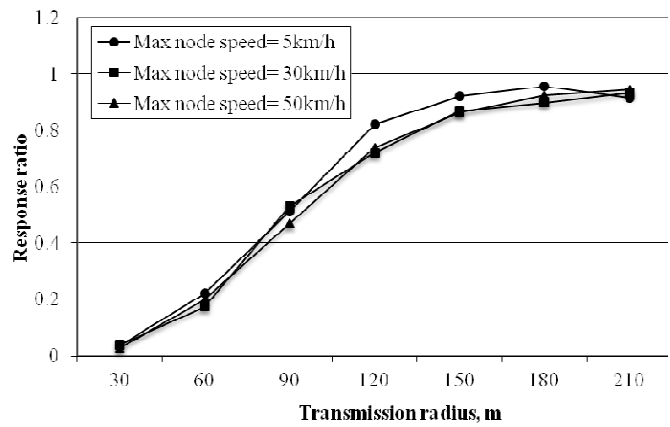

Figure 2: Response ratio versus transmission radius with link availability *l*=0.7 and different maximal node speed.

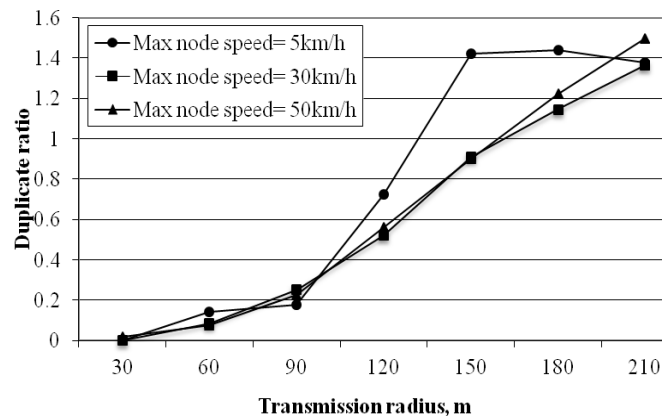

Figure 3: Duplicate ratio versus transmission radius with link availability *l*=0.7 and different maximal node speed.

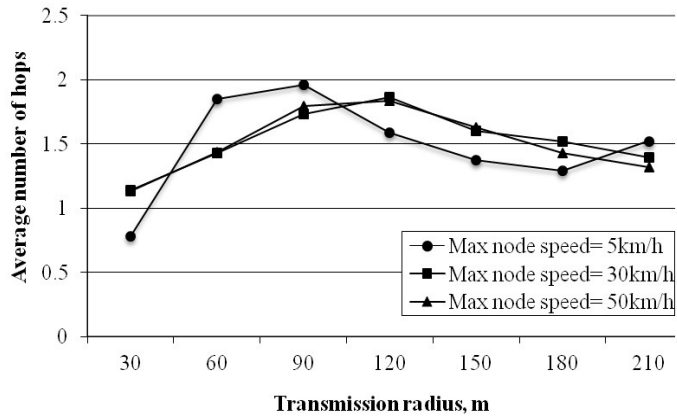

Figure 4: Average number of hops versus transmission radius with link availability *l*=0.7 and different maximal node speed.

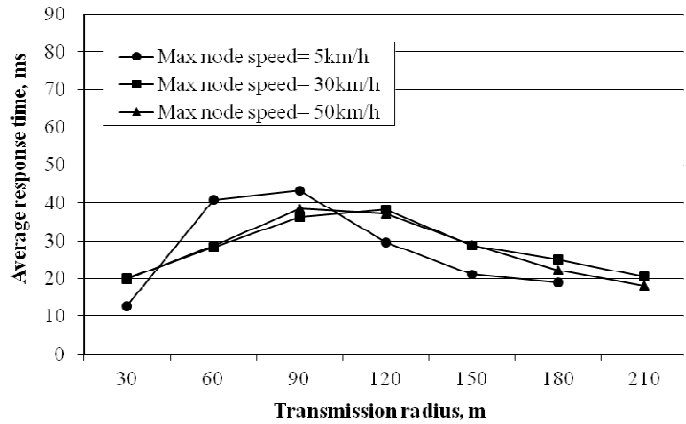

Figure 5: Average response time versus transmission radius with link availability *l*=0.7 and different maximal node speed.

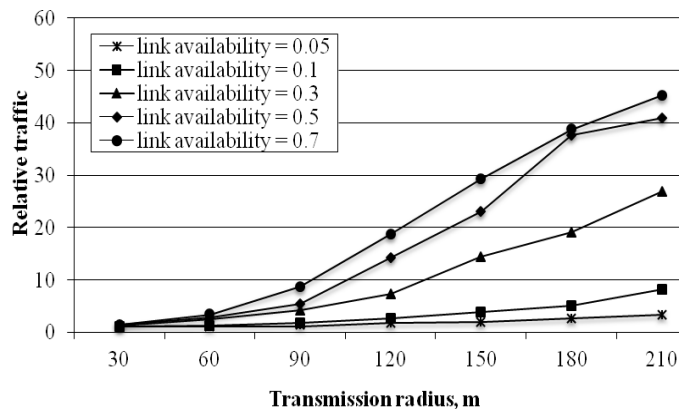

Figure 6: Relative traffic versus transmission radius with different link availability for TTL=4.

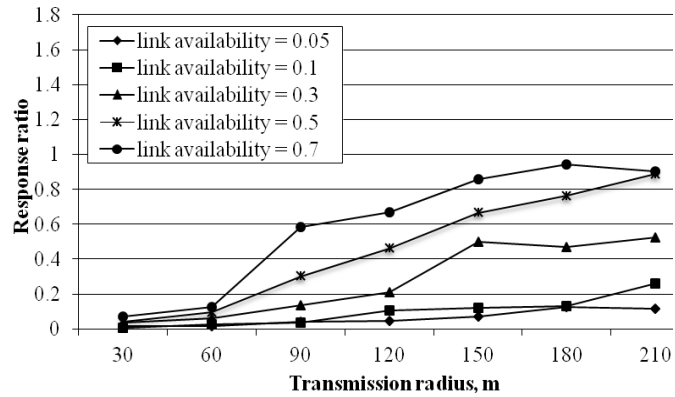

Figure 7: Response ratio versus transmission radius with different link availability for TTL=4.

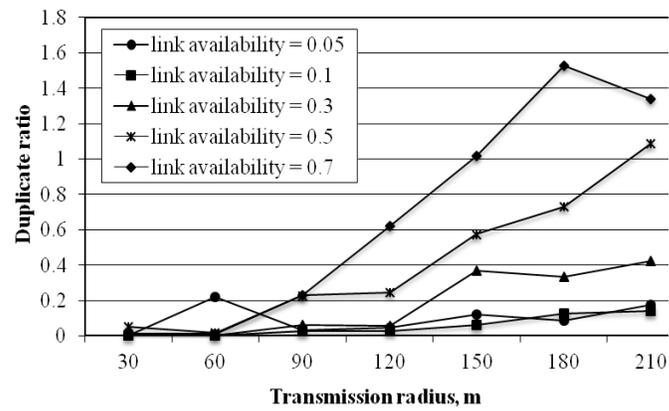

Figure 8: Duplicate ratio versus transmission radius with different link availability for TTL=4.

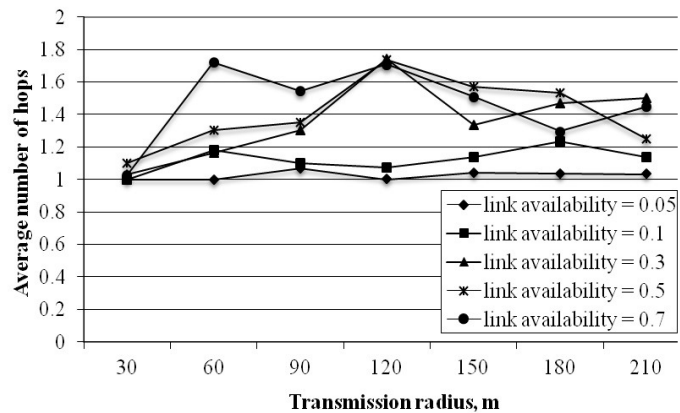

Figure 9: Average number of hops versus transmission radius with different link availability for TTL=4.

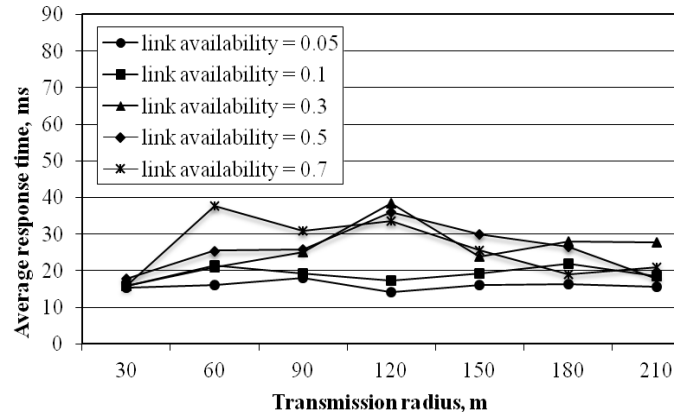

Figure 10: Average response time versus transmission radius with different link availability for TTL=4.

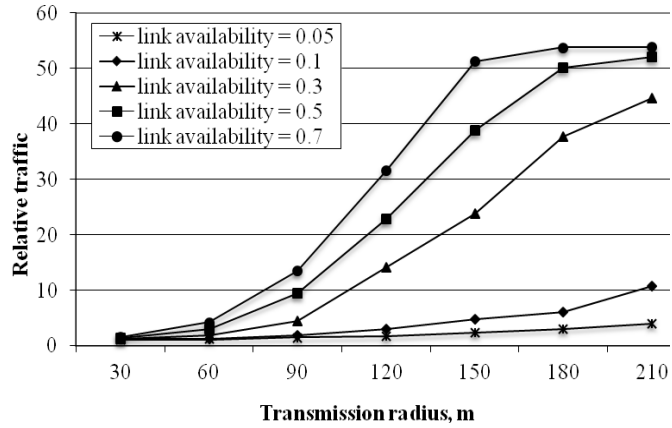

Figure 11: Relative traffic versus transmission radius with different link availability for TTL=7.

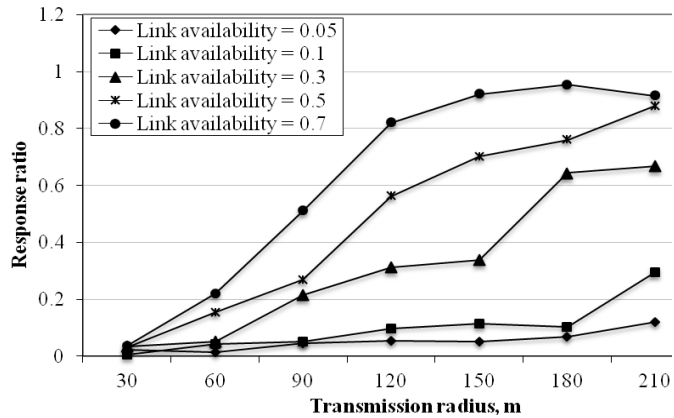

Figure 12: Response ratio versus transmission radius with different link availability for TTL=7.

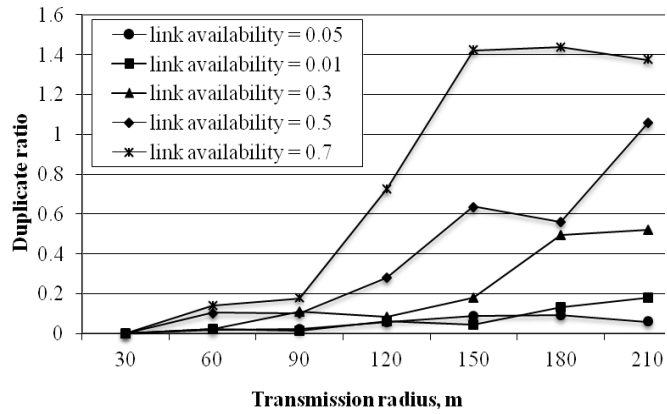

Figure 13: Duplicate ratio versus transmission radius with different link availability for TTL=7.

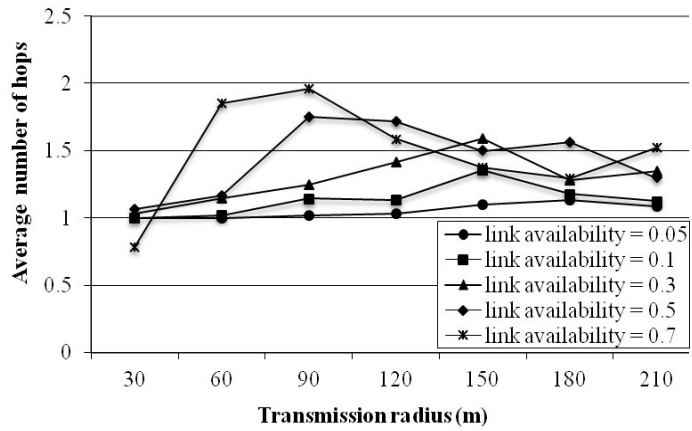

Figure 14: Average number of hops versus transmission radius with different link availability for TTL=7.

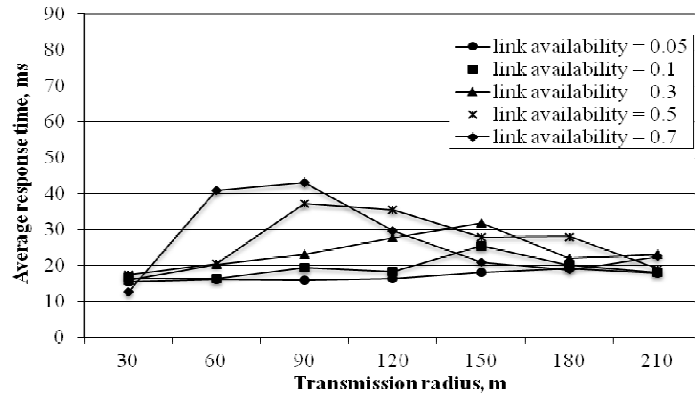

Figure 15: Average response time versus transmission radius with different link availability for TTL=7.

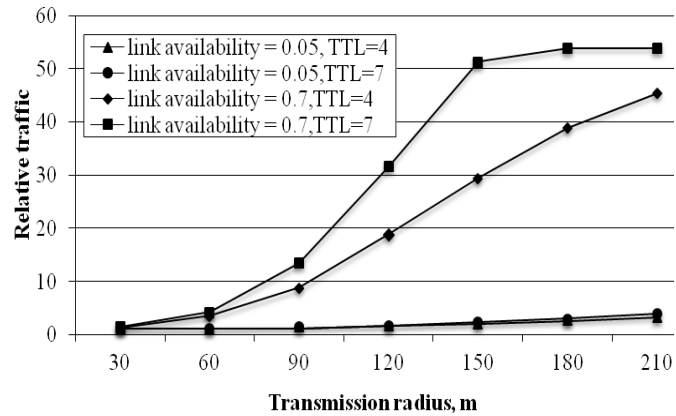

Figure 16: Relative traffic versus transmission radius with different link availability for TTL=7, and 4.

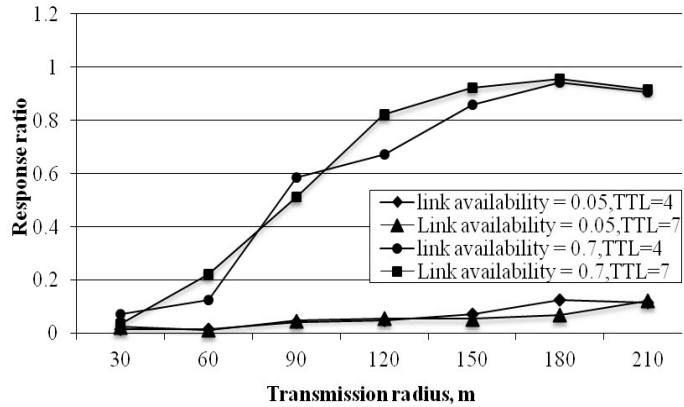

Figure 17: Response ratio versus transmission radius with different link availability for TTL=7 and 4.

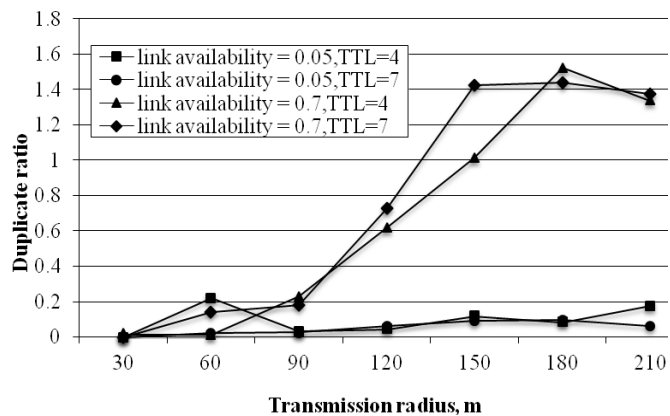

Figure 18: Duplicate ratio versus transmission radius with different link availability for TTL=7 and 4.

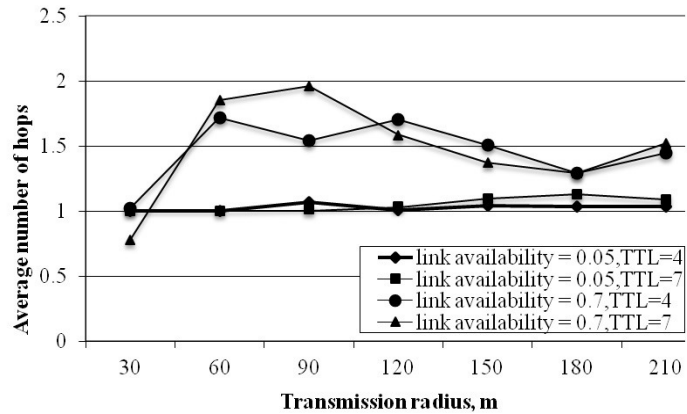

Figure 19: Average number of hops versus transmission radius with different link availability for TTL=7 and 4.

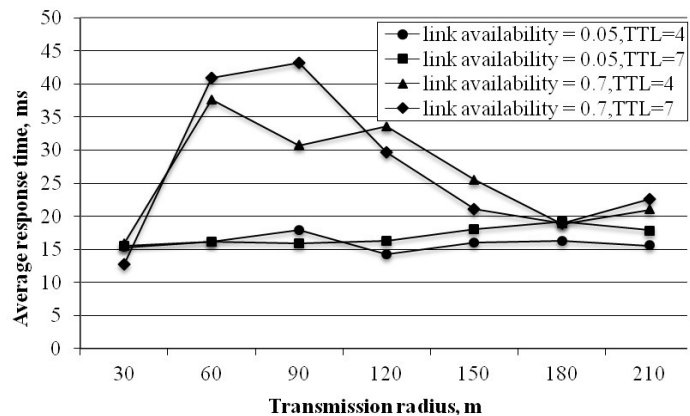

Figure 20: Average response time versus transmission radius with different link availability for TTL=7 and 4.

## 5. DISCUSSIONS OF THE RESULTS

The following comments and observations can be made using the simulation results:

1. All the performance metrics dependent on the transmission radius, but the character of this dependence is different for different performance metrics.

2. As Figures 2, 7 and 12 demonstrate, the response ratio is quite low for small values of transmission radius, but it approaches the highest value of 1 at the transmission radius of 210m. However, for small link availability l=0.05, the response ratio remains quite low even at transmission radius of 210m, since a large number of packets are lost on the path from the source node to server and back.

3. At a small transmission radius of 30 m, the response ratio is low even for high value of link availability l = 0.7. The reason is that, with N = 50 nodes in the network, there is a high probability that each transmitting or forwarding node has no neighbors within this transmission radius. This means that a packet transmitted by a node in network area has a very low chance to be received by at least one other node in this area.

4. Response ratio has direct relation with link availability. Increasing the link availability results in increment of response ratio, and it becomes more obvious in higher transmission radiuses (Figures 7 and 12). Also, as you can understand from Figures 2, and 17 different node speeds, and TTL has no significant effect on this performance metric.

5. As Figures 4, 9 and 14 show, the average number of hops is quite low at a small transmission radius. It initially increases with the increase of the transmission radius, reaching some maximum and then decreases. Such a behavior of this metric can be explained in the following way. When the transmission radius is small, then, as it was explained earlier, many transmitted or forwarded packets will be received mainly by a close neighbor. It means, the packet can reach the destinations if only the destination is a close neighbor of source, with a low number of hops. On the other hand, with a very large transmission radius, many nodes will find their destination node in the coverage area, so packet can be transmitted with only one transmission. This reduces the average number of hops again.

6. The average number of hops metric was usually varying in the range (1-2) for different link availabilities and node speeds. Figures 9 and 14 indicate that for small link availabilities (l=0.05 and 0.1) it shows small changes and always stays close to 1 even with increasing the transmission radius. This performance metric is the same for different TTL with small link availability as shown in Figure 19, but it reaches higher average number of hops for TTL=7. As packets with smaller TTL will be discarded on their way to destination, it results in a minor increase in this performance metric.

7. The third metric, the relative traffic, can be quite high for a large value of link availability (Figures 6, and 11), especially at large transmission radius, when more and more nodes are involved in the retransmission of packets (Figure 1). With variable values of TTL, the number of nodes involved in packet transmission is reduced. As shown in Figure 16, a value of TTL=4 has a small impact on the performance of the pure flooding scheme.

8. As Figures 5, 10 and 15 show, the average response time is quite low at a small transmission radius. It initially increases with the increase of the transmission radius, reaches some maximum and then decreases. As explained before, when the transmission radius is small, less numbers of nodes are involved in the transmission. On the other hand, with a large transmission radius, many transmitted packets will find their destination node in the area with only one transmission. This reduces response time of the packets. Plus, maximum average response time is larger for packets with bigger TTL as you can see in Figure 20.

9. Figures 3, 8, and 13 show that the duplicate ratio, the last metric, is quite low for a small transmission radius, but can be high for a large value of link availability (Figures 8, and 13). At the transmission radius of 210m as it approaches the highest value, since more than one server can be in the range of the transmitted packets and contribute to duplicate replies.

10. As indicated in Figure 18, in a network with all its nodes having the same link availability, changing the packet's TTL doesn't have a visible effect on the duplicate ratio.

11. As graphs in Figures 1-5 demonstrate, change of the maximum possible node speed in the range from 5 Km/h to the medium speed of a car in a city of 50 Km/h does not result in considerable change of all performance metrics.

## 6. CONCLUSIONS

In this work, an anycast flooding simulation model of mobile ad hoc networks was used to investigate practically important performance metrics. The dependence of practically important performance metrics on the transmission radius, link availability and maximal possible node speed was investigated by conducting a large number of simulation studies. These metrics are the delivery ratio, the average number of hops, the relative traffic, the response time and the duplicate ratio.

Simulation results show that changing the maximum node speed has small effect on performance of the network. For the small link availabilities performance metrics remain quite low even at transmission radius of 210m. On the other hand, decreasing the TTL can result in less traffic when other performance metrics are nearly same.

In summary, the present simulation results together with the simulation model can be used for better and closer understanding toward anycasting in ad hoc wireless networks.

**Author**

**Reza Azizi** was born in Bojnourd, Iran in 1983. He received the BS degree in Software Engineering from Islamic Azad University of Shirvan, Iran, a MS degree in Computer Engineering from Eastern Mediterranean University, Cyprus. He has been employed as a research assistant at computer engineering department, where he has been involved in multiple A class projects. He is currently working as Lecturer at Islamic Azad University of Bojnourd and a joint member of Educational Center of Applied-Science and Technology. Research interests: Simulation, Performance Evaluation and Optimization of Wireless Sensor and Mobile Ad-Hoc Networks.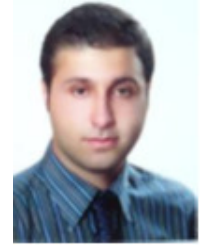